\documentclass[twocolumn,showpacs,amsmath,amssymb,prd,nofootinbib,
superscriptaddress]{revtex4-1}

\usepackage{dcolumn}
\usepackage{bm}
\usepackage{ifpdf}
\usepackage{hyperref}
\usepackage{dcolumn}
\usepackage{bm}
\usepackage[spanish,english]{babel}
\usepackage{amsfonts}
\usepackage{amssymb}
\usepackage{graphicx}
\usepackage{color}
\usepackage[latin1]{inputenc}
\newcommand{\LL}{\mathcal{L}}

\newcommand \be{\begin{equation}}
\newcommand \en{\end{equation}}
\newcommand \bea{\begin{eqnarray}}
\newcommand \ena{\end{eqnarray}}

\begin{document}

\title{Palatini wormholes and energy conditions from the prism of General Relativity}

\author{Cecilia Bejarano} \email{cbejarano@iafe.uba.ar}
\affiliation{Instituto de Astronom\'ia y F\'isica del Espacio (IAFE, CONICET-UBA),
Casilla de Correo 67, Sucursal 28, 1428 Buenos Aires, Argentina.}
\affiliation{Departamento de F\'{i}sica Te\'{o}rica and IFIC, Centro Mixto Universidad de
Valencia - CSIC. Universidad de Valencia, Burjassot-46100, Valencia, Spain.}

\author{Francisco S. N. Lobo}
\email{fslobo@fc.ul.pt}
\affiliation{Instituto de Astrof\'{\i}sica e Ci\^{e}ncias do Espa\c{c}o, Faculdade de
Ci\^encias da Universidade de Lisboa, Edif\'{\i}cio C8, Campo Grande,
P-1749-016 Lisbon, Portugal.}

\author{Gonzalo J. Olmo} \email{gonzalo.olmo@uv.es}
\affiliation{Departamento de F\'{i}sica Te\'{o}rica and IFIC, Centro Mixto Universidad de
Valencia - CSIC. Universidad de Valencia, Burjassot-46100, Valencia, Spain.}
\affiliation{Departamento de F\'isica, Universidade Federal da
Para\'\i ba, 58051-900 Jo\~ao Pessoa, Para\'\i ba, Brazil.}

\author{Diego Rubiera-Garcia} \email{drgarcia@fc.ul.pt}
\affiliation{Instituto de Astrof\'{\i}sica e Ci\^{e}ncias do Espa\c{c}o, Faculdade de
Ci\^encias da Universidade de Lisboa, Edif\'{\i}cio C8, Campo Grande,
P-1749-016 Lisbon, Portugal.}


\date{\today}

\begin{abstract}
Wormholes are hypothetical shortcuts in spacetime that in General Relativity unavoidably violate all of the pointwise energy conditions. In this paper, we consider several wormhole spacetimes that, as opposed to the standard \emph{designer} procedure frequently employed in the literature, arise directly from gravitational actions including additional terms resulting from contractions of the Ricci tensor with the metric, and which are formulated assuming independence between metric and connection (Palatini approach). We reinterpret such wormhole solutions under the prism of General Relativity and study the matter sources that thread them. We discuss the size of violation of the energy conditions in different cases, and how this is related to the same spacetimes when viewed from the modified gravity side.
\end{abstract}

\maketitle

\section{Introduction}

There exist numerous exact solutions of the Einstein field equations in which the physical relevance is not always clear and, consequently, are often regarded as exotic \cite{Mc}. For instance, a well known example is the static and spherically symmetric Schwarzschild solution, where the importance of this geometry was not fully understood until the observational discovery of very compact objects such as neutron stars. In the 1960s, the Schwarzschild solution began to be considered seriously not only as an approximate description of the exterior geometry of (non-rotating) stars but also as the final outcome of gravitational collapse \cite{Joshi}. Nowadays, black holes are a hot topic of research, such as the sources of gravitational waves  \cite{TheLIGOScientific:2016wfe}, and are no longer regarded as exotic solutions.

Wormholes are also regarded as an exotic family of solutions of Einstein's equations \cite{mortho88,visser,Lobo:2007zb}, and have been extensively analyzed for a number of reasons. From a quantum gravitational perspective, they are seen as a natural consequence if the topology of spacetime fluctuates in time \cite{whe55}. Recently, a new connection between specific types of (non-traversable) wormholes \cite{eisros35} and quantum systems was introduced with relevant implications in high energy physics \cite{epr-er}. From an astrophysical and cosmological viewpoint, the late-time cosmic speedup \cite{obs-data} suggests that the source driving  the expansion \cite{cosmo} could be compatible with the kind of matter necessary to generate traversable wormholes. In fact, several cosmological models with exotic sources supporting wormholes have been proposed, including phantom energy \cite{phantom}, Chaplygin gas \cite{chaply}  or its generalizations \cite{gen-chaply}. This has led many researchers to explore the astrophysical consequences and characterization of macroscopic wormholes \cite{astroWH}.

In the context of General Relativity (GR) wormhole spacetimes are specifically designed, where Einstein's equations are solved in a reversed manner. This means that a suitable well behaved geometry is defined {\it a priori} and the matter sources that generate it are derived {\it  a posteriori}. The outstanding feature was the requirement of {\it exotic matter}, defined as matter that violates the null energy condition, i.e., $T_{\mu\nu} k^{\mu}k^{\nu}<0$, where $T_{\mu\nu}$ is the stress-energy tensor and $k^{\mu}$ is any null vector \cite{mortho88, visser, hocvis}. Although exotic matter is considered classically non-viable, there are experimental evidences of such energy conditions violations in several systems where quantum effects such as the Casimir effect, Hawking evaporation or vacuum polarization take place \cite{visser,vis96}. Therefore, since the existence of exotic matter is classically a problematic issue, it is particularly important to find solutions which minimize the violation of the energy conditions. In fact, the amount of violation can be made infinitesimally small by choosing the geometry of the wormhole in a very specific and appropriate way \cite{viskardad03}. The thin-shell formalism is another approach to minimize the exotic matter, as the latter is now localized on the thin-shell \cite{vis89,thw-sphsym}.

However, in the context of modified gravity, it was shown explicitly that the matter threading the wormhole throat may, in principle, satisfy all of the energy conditions, and it is the higher order curvature terms, interpreted as a gravitational fluid, that support the wormhole geometry \cite{harko-etal13}. In fact, a plethora of wormhole solutions beyond GR have been found in a variety of theories, such as in conformal Weyl gravity \cite{weylgrav}, Kaluza-Klein \cite{kk}, Brans-Dicke \cite{bd}, Gauss-Bonnet \cite{gb}, Lovelock \cite{lovelock}, braneworlds \cite{bw}, Ho\u{r}ava-Lifshitz \cite{hl}, Eddington-inspired Born-Infeld \cite{bi}, the metric \cite{f(R)} and Palatini \cite{bcor15,or15} approaches of $f(R)$ gravity and extensions \cite{nonmin}, modified teleparalellism framework \cite{tele-ext} or Einstein-Cartan theory \cite{ec}, amongst others.

The purpose of this work is to consider specific wormhole solutions that can be obtained in gravitational theories formulated in the Palatini approach, and interpret them from the prism of standard GR. This means that considering specific wormhole solutions, and taking into account the modified Einstein's equation, it is then possible to interpret its associated effective stress-energy tensor from a standard perspective and determine if it violates the generalized energy conditions \cite{harko-etal13}. In this sense, the relevance and novelty of this work lies in the fact that the wormhole solutions considered here do not arise as a result of the \emph{reversed philosophy} procedure, where the space-time geometry is given first and then the Einstein's equations are driven back to find the matter source threading such a geometry, but instead flows from the resolution of the modified Einstein's equations associated to a well-defined gravitational actions without ghosts (see Sec. \ref{sec:IV} and \ref{sec:V} below). More specifically, a thorough analysis of the specific solutions dictates that these wormhole solutions are inherent in the Palatini theories considered in this work. Thus this approach allows one to study wormhole geometries in less artificial scenarios and to investigate such geometries from the GR point of view in relation to violation of the energy conditions.

The paper is organized as follows. We begin by introducing the basics of Palatini theories in Sec. \ref{sec:I}. In Sec. \ref{sec:II}, we briefly review wormholes physics in GR. In Sec. \ref{sec:IV}, we present specific Palatini wormholes, namely, wormholes obtained in Born-Infeld gravity formulated in the Palatini approach, and discuss them under the prism of GR, identifying the matter source threading the wormhole and discussing the violation of the energy conditions. In Sec. \ref{sec:V}, this analysis is extended to wormholes obtained in the context of $f(R)$ gravities, and we conclude in Sec. \ref{sec:VI} with a summary and discussion.

\section{Palatini theories of gravity} \label{sec:I}

The wormhole solutions studied here have been found in theories of gravity formulated \`{a} la Palatini, where one assumes that the metric and the connection are independent geometric entities. In four spacetime dimensions, an affine connection $\Gamma^\lambda_{\mu\nu}$ may have up to 64 independent components. Its antisymmetric part is known as the torsion tensor $S^\lambda_{\mu\nu}=(\Gamma^\lambda_{\mu\nu}-\Gamma^\lambda_{\nu\mu})/2$, which represents 24 degrees of freedom, while its symmetric part $C^\lambda_{\mu\nu}=(\Gamma^\lambda_{\mu\nu}+\Gamma^\lambda_{\nu\mu})/2$ represents the remaining 40 degrees of freedom. When a metric is considered, the nonmetricity tensor $Q_{\alpha\mu\nu}=\nabla^{\Gamma}_\alpha g_{\mu\nu}$ can be used together with the torsion and the metric to fully determine the symmetric part of the connection as
\begin{eqnarray}\label{eq:C}
C^\lambda_{\mu\nu}&=&L^\lambda_{\mu\nu}+g^{\lambda\alpha}g_{\rho\nu}S^\rho_{\mu\alpha}
+g^{\lambda\alpha}g_{\rho\mu}S^\rho_{\nu\alpha}
	\nonumber \\
&& + \frac{g^{\lambda\alpha}}{2}\left(Q_{\mu\alpha\nu}+Q_{\nu\alpha\mu}-Q_{\alpha\mu\nu}\right) \ ,
\end{eqnarray}
where $L^\lambda_{\mu\nu}$ represents the Levi-Civita connection of the metric, namely
\begin{equation}
L^\lambda_{\mu\nu}= \frac{g^{\lambda\alpha}}{2}\left(\partial_\mu g_{\alpha\nu}+\partial_\nu g_{\alpha\mu}-\partial_\alpha g_{\mu\nu}\right) \ .
\end{equation}
We note that most of the literature on metric-affine gravity has focused on Einstein-Cartan spaces (with torsion but no nonmetricity \cite{EC}). Here, we consider torsionless scenarios with nonmetricity, which naturally arises when one considers gravity Lagrangians beyond GR. For simplicity, we will assume that the matter action only depends on the metric, which is compatible with the experimental evidence supporting the Einstein Equivalence Principle \cite{Willreviews}. Note, however, that the matter action could depend on the connection in regimes not yet explored experimentally. In fact, since the coupling between fermions and gravity is mediated by the torsion \cite{Ortin}, this possibility should be considered at some stage. Here, we follow a conservative approach in this respect and will restrict ourselves to spinless matter fields.

\subsection{$f(R)$ theories} \label{eq:secIIB}

Consider extensions of Einstein's theory in the $f(R)$ Palatini approach, where we will omit details about the derivation of the field equations, as they are available in the literature \cite{olmorev}. The variation with respect to the metric and the connection lead to
\begin{eqnarray}
f_R R_{\mu\nu}(\Gamma)-\frac{f}{2} g_{\mu\nu}&=&\kappa^2 T_{\mu\nu} ,
\end{eqnarray}
\begin{eqnarray}
\nabla_\alpha \left[\sqrt{-g}f_R g^{\mu\nu}\right]&=&0 \ , \label{eq:conn-var}
\end{eqnarray}
respectively, where $f_{R}\equiv df/dR$, while $T_{\mu\nu}$ is the stress-energy tensor of the matter, which is assumed to couple minimally to gravity. Tracing the first of these equations with the metric $g^{\mu\nu}$ one obtains: $R f_R-2f=\kappa^2 T$.
This is an algebraic equation that relates $R$ with the trace of the matter stress-energy tensor, $T$. In the case of GR, that relation is linear, $R=-\kappa^2T$, but for nonlinear Lagrangians the relation will be, in general, nonlinear. In any case, note that $R$ can be written as $R=R(T)$, implying that any function of $R$ will be a function of $T$. As a result, the connection equation (\ref{eq:conn-var}) can be seen as algebraic and linear for $\Gamma^\alpha_{\mu\nu}$ because the $R$ dependences can be traded by functions of $T$. One thus finds that introducing an auxiliary metric, $h_{\mu\nu}=f_R g_{\mu\nu}$, Eq.~(\ref{eq:conn-var}) turns into $\nabla_\alpha \left[\sqrt{-h}h^{\mu\nu}\right]=0$. This equation is equivalent to $\nabla_\alpha h_{\mu\nu}=0$, whose solution is the well known Levi-Civita connection of $h_{\mu\nu}$
 \begin{equation}
\Gamma^\lambda_{\mu\nu}= \frac{h^{\lambda\alpha}}{2}\left(\partial_\mu h_{\alpha\nu}+\partial_\nu h_{\alpha\mu}-\partial_\alpha h_{\mu\nu}\right) \ .
\end{equation}
By writing explicitly the dependence on $g_{\mu\nu}$ in this equation, one obtains the result advanced in Eq. (\ref{eq:C}) with $Q_{\alpha\mu\nu}=\nabla^{\Gamma}_\alpha g_{\mu\nu}=-(\partial_\alpha \ln f_R) g_{\mu\nu}$.  The metric field equations can then be written in compact form as
\begin{equation}\label{eq:Rmnh}
{R^\mu}_\nu(h)=\frac{1}{f_R^2}\left(\frac{f}{2}\delta^\mu_\nu+\kappa^2{T^\mu}_\nu\right) \ ,
\end{equation}
where ${R^\mu}_\nu(h)=h^{\mu\alpha}R_{\alpha\nu}(h)$ and $R_{\alpha\nu}(h) \equiv R_{\alpha\nu}(\Gamma)$. Note that the $R-$dependent functions on the right-hand side of this equation are functions of the matter via the trace $T$. With these equations, one could solve for $h_{\mu\nu}$ once the matter sources are specified and then use the relation  $h_{\mu\nu}=f_R g_{\mu\nu}$ to obtain $g_{\mu\nu}$. This strategy, however, is not always straightforward. One could thus opt in writing  ${R^\mu}_\nu(h)$ in terms of ${R^\mu}_\nu(g)$ and get a set of equations directly referred to $g_{\mu\nu}$. Thus, Eq. (\ref{eq:Rmnh}) turns into
\begin{eqnarray}\label{eq:Gmn_f(R)}
G_{\mu\nu}(g)&=&\kappa^2\tau_{\mu\nu}\ ,
\end{eqnarray}
where the modified (effective) stress-energy tensor, $\tau_{\mu\nu}$, is defined as
\begin{eqnarray}\label{eq:Tmn_MOD}
\kappa^2\tau_{\mu\nu}&=&\frac{\kappa^2}{f_R}T_{\mu\nu}-\frac{Rf_R-f}{2f_R}g_{\mu\nu}
	\nonumber \\
&&- \frac{3}{2f_R^2}\left[\partial_\mu f_R\partial_\nu f_R-\frac{1}{2}g_{\mu\nu}(\partial f_R)^2\right]
	\nonumber\\ &&+\frac{1}{f_R}\left[\nabla_\mu \nabla_\nu f_R-g_{\mu\nu}\Box f_R\right] \ .
\end{eqnarray}
Written in this form, these equations can be interpreted as in Einstein's theory but with the right-hand side of Eq. (\ref{eq:Gmn_f(R)}) containing the stress-energy tensor  $T_{\mu\nu}$ plus other functions which depend on the trace of the matter via $R(T)$. It is precisely this form of the equations that we will use to reinterpret $f(R)$ wormhole solutions as solutions of Einstein's theory with a modified matter source. In other words, it is important to bear in mind that the modified stress-energy tensor $\tau_{\mu\nu}$ contains extra contributions which are just functions of the matter.

\subsection{Beyond the $f(R)$ case} \label{sec:IIB}

It has been recently established \cite{olmobook} that Palatini theories in which the gravity Lagrangian is a function of the Ricci tensor and the metric, $\LL_G=f[g_{\alpha\beta}, R_{\mu\nu}(\Gamma)]$, lead to field equations of the form
\begin{equation}\label{eq:Rmn}
{R^\mu}_\nu(h)=\frac{\kappa^2}{|\hat\Omega|^{1/2}}\left(\LL_G\delta^\mu_\nu+{T^\mu}_\nu\right) \ ,
\end{equation}
where $|\hat\Omega|$ represents the determinant of a matrix which relates the physical metric $g_{\mu\nu}$ and an auxiliary metric $h_{\mu\nu}$ which is compatible with the affine connection that defines the Ricci tensor. In other words, $h_{\mu\nu}=\Omega_{\mu}{}^\alpha g_{\alpha\nu}$ and $\nabla^\Gamma_\alpha h_{\mu\nu}=0$.

In these theories, the nonmetricity tensor takes the form $Q_{\alpha\mu\nu}=g_{\rho\nu}(\nabla_\alpha {{\Omega^{-1}}_{\mu}}^\lambda) {\Omega_\lambda}^\rho $ (symmetrized over $\mu$ and $\nu$), where ${\Omega_{\mu}}^\alpha$ is a function of $T_{\mu\nu}$. In the $f(R)$ case, one has that ${\Omega_{\mu}}^\alpha=f_R \delta_\mu^\alpha$ (conformal transformation), which leads to a simple nonmetricity tensor. In general, however, the relation between the metrics is not conformal and a {\it clean} representation of the field equations in terms of $g_{\mu\nu}$ is not immediate. For this reason, Eq.~(\ref{eq:Rmn}) is typically preferred. Nonetheless, for the spherically symmetric scenarios that will be considered here, the Einstein tensor $G_{\mu\nu}(g)$ can always be computed straightforwardly, which facilitates the interpretation of the solutions in terms of an effective $\tau_{\mu\nu}$ in much the same way as in $f(R)$ theories.

As the field equations can be written in Einstein's-like form (\ref{eq:Gmn_f(R)}), a given space-time (wormhole or not) can be interpreted as a solution of an (in principle) infinitely degenerate family of theories of gravity. Here, we will show that these wormhole solutions are not designed, but follow directly from well motivated actions. This could shed light on obtaining analytic wormhole solutions and on the kind of energy sources that can generate them in the context of GR. The reinterpretation  of these wormhole solutions from the prism of GR might be useful to understand their properties from a more standard perspective. To proceed, we review wormholes physics in next section.

\section{General theory of wormholes} \label{sec:II}

In 1955, John Wheeler introduced the term {\it geon} to denote a hypothetical gravitational-electromagnetic object with a nontrivial topological structure \cite{whe55}, and later with Misner \cite{miswhe57} coined such a construction a \emph{wormhole}. These concepts extended the very first calculations tracing back to the Einstein and Rosen bridge \cite{eisros35}.  But it was the seminal work of Morris and Thorne \cite{mortho88} in 1988 which gave rise to the great interest on traversable Lorentzian wormholes which continues today. In this section, such concepts will be introduced, and the transition from wormholes in GR to those in metric-affine geometries will become clear.

\subsection{Wormhole geometry} \label{sec:IIA}

Lorentzian wormholes are nontrivial topological structures connecting two asymptotically flat regions, either in the same universe, or in different ones. The geometry of a static and spherically symmetric traversable wormhole spacetime can be written as \cite{mortho88,visser}
\begin{equation} \label{eq:whm2}
ds^2= -e^{2\Phi(r)} dt^2 + \frac{1}{1-b(r)/r} dr^2 + r^2 d\Omega^2 \ ,
\end{equation}
where $(t,r,\theta,\phi)$ are the standard Schwarzschild coordinates, and $d\Omega^2=d\theta^2 + \sin^2 \theta d \varphi^2$ denotes the spherical sector. $\Phi(r)$ is called the redshift function because it is related to the gravitational redshift, and $b(r)$ is denoted the shape function as it determines the shape of the wormhole \cite{mortho88}. In order to describe the wormhole solution, the radial coordinate possesses a minimum at $r_0$, which defines the wormhole throat.
The latter connects two asymptotically flat regions and it is defined, for static configurations, as a surface of minimum area that satisfies the flaring-out condition, which can be obtained from the embedding calculations for the wormhole geometry and reads $[b(r)-b'(r)r]/b^{2}(r) > 0$. One thus deduces that $b'(r_0) < 1$ must hold, since at the throat $b(r_0)=r_0$ is imposed.

The functions $\Phi(r)$ and $b(r)$ fulfil further properties in order to describe a wormhole spacetime. The asymptotically flat limit imposes that $b(r)/r \rightarrow 0$, as $r \rightarrow \infty$. Moreover, for the wormhole to be traversable there should be no event horizons so that $\Phi(r)$ must be finite everywhere. Note that it is straightforward to define the mass of the wormhole as the finite limit for the function $b(r)$. By comparison with the Schwarzschild geometry, an observer from spatial infinity sees that $b=2G_NM$.

The line element of a wormhole geometry can be alternatively written as
\begin{equation} \label{eq:whm1}
ds^2=-e^{2\Phi(x)} dt^2 + dx^2 + r^2(x)d\Omega^2 \ ,
\end{equation}
where the radial coordinate $x$ is interpreted as the proper distance from the throat, located at $x=0$, and related to $r$ by $dx/dr=\pm 1/\sqrt{1-b(r)/r}$. The advantage of this new set of coordinates lies in that it allows one to describe the wormhole geometry by using a unique chart with $x \in (-\infty,+\infty)$. In terms of $x$, then it is clear that the area of the spherical surfaces never drops below $4\pi r_0^2$.

A useful way to construct wormhole spacetimes resides in the cut-and-paste procedure, which is a mathematical construction where two spacetimes are matched at a given junction interface. If this hypersurface contains non-zero surface stresses it is called a {\it thin-shell}, otherwise it is just a boundary surface. Following the standard junction-condition formalism \cite{:junction:}, one cuts and pastes two manifolds to form a geodesically complete one with the throat located at the joining shell, where the exotic matter is located \cite{vis89,thw-sphsym}. Beyond GR, the junction formalism requires to be generalized for the specific theory of gravity under consideration, for example, such as in metric $f(R)$ gravity \cite{junction-f(R)}.

At this point we would like to point out a special feature of the thin-shell structure. It is well known that thin-shell wormholes are geometric constructions that turn out to be geodesically complete although the Riemann tensor is divergent at the thin-shell where the throat is located \cite{poisson}. Thus, the spacetime curvature becomes divergent at the non-null hypersurface layer, however, this divergence is physically interpreted as a surface layer with a stress-energy tensor on it. Therefore, the existence of curvature divergences at the wormhole throat should not be surprising at all, and this does not necessarily entail the presence of spacetime singularities. Note, in this sense, that we are using the term {\it singularity} in a way that transcends the notion of {\it divergence}. A spacetime is said to be {\it singular} when it is geodesically incomplete \cite{Wald, Earman}, regardless of the existence or not of curvature divergences. Thus, these terms will not be interchangeable in our discussion. In fact, thin-shell wormholes are examples of spacetimes which are geodesically complete (hence nonsingular) but which, by construction, contain curvature divergences.

\subsection{Energy conditions} \label{ec-gr}

Since the foundation of GR, a number of standard energy conditions \cite{visser}, which are assumptions about the energy-matter content representing physically realistic situations found in Nature, have been thoroughly studied and classified in the literature. Assuming a diagonal stress-energy tensor of the form
\begin{equation}\label{SETdef}
T^{\mu}{}_{\nu}={\rm diag}(-\rho,p_1,p_2,p_3)\,,
\end{equation}
these conditions read:

\begin{itemize}
  \item The weak energy condition (WEC) states that the energy density measured by an arbitrary observer must be non-negative, $\rho \geq 0$ and  $\rho + p_{i} > 0$.
  \item The strong energy condition (SEC) asserts that gravity is attractive, $\rho + \sum p_{i}\geq 0$ and $\rho + p_{i} \geq 0$. Note that this condition is violated in many current models of accelerated cosmic expansion as well as in inflationary models.
  \item The dominant energy condition (DEC) expresses that the energy density measured by any observer is positive but also that its flux propagates in a causal way (i.e., it is null or timelike)  so that $\rho \geq 0$ and $\rho \geq |p_{i}|$.
  \item The null energy condition (NEC) implies that $\rho + p_{i} \geq 0$, $i=1,2,3$.
\end{itemize}

Since WEC or SEC $\Rightarrow$ NEC and DEC $\Rightarrow$ WEC $\Rightarrow$ NEC,  all energy conditions are automatically violated if the NEC is not satisfied. In particular, due to the flaring-out condition, wormhole spacetimes violate the NEC at the throat itself, and consequently violate all of the above pointwise energy conditions. Quantum effects can also produce violations of the classical energy conditions, which in order to support microscopic wormholes must typically be of order $\sim \hbar$. It is worth mentioning that there exist averaged versions of the energy conditions since quantum violations are probably not allowed globally (see \cite{visser} for more details).

Note that the energy conditions in GR can be traced back to the Raychaudhuri equation, where it is straightforward to determine that the  attractive nature of gravity is represented by the condition that the Ricci tensor fulfils $R_{\mu \nu} k^\mu k^\nu \geq 0$ for any null vector $k^{\mu}$. This condition ensures the focusing of the geodesic congruence,  which in turn can be written, via Einstein's field equations, as a condition over the stress-energy tensor $T_{\mu \nu} k^\mu k^\nu \geq 0$ \cite{hawking-ellis}. On the contrary, the wormhole structure requires that the null geodesic congruence must be defocused at the throat in order for geodesic completeness to hold. This is precisely the physical meaning encoded in the flaring-out condition.

In extended theories of gravity, it is possible to express the gravitational field equations as in Eq. (\ref{eq:Gmn_f(R)}) where the effective stress-energy tensor $\tau_{\mu \nu}$ includes all new theory-dependent terms as well as the corresponding stress-energy tensor of the matter \cite{Capozziello:2013vna}. Thus, if one has repulsive gravity $R_{\mu \nu} k^\mu k^\nu < 0$, this implies $\tau_{\mu \nu} k^\mu k^\nu < 0$, but the matter stress-energy tensor can, in principle, be imposed to obey the energy conditions, or more specifically, in this case the NEC, i.e., $T_{\mu \nu} k^\mu k^\nu \geq  0$. Thus, in the context of modified gravity it has been shown that wormhole geometries are in fact supported by the effective stress-energy tensor, which actually plays the role of exotic matter and violates the energy conditions, while the physical matter satisfies them \cite{harko-etal13}.

\section{Wormholes in Born-Infeld and quadratic Palatini gravity} \label{sec:IV}

We will now study a family of wormholes which arises as an exact solution of an extension of GR containing quadratic curvature scalar and Ricci-squared terms \cite{or}, and also of Born-Infeld gravity \cite{ors}, a theory that has attracted a good deal of attention in astrophysics and cosmology in the last few years \cite{BIg} (see \cite{BHOR17a} for a recent review). Quadratic extensions of GR are supported by the quantization of fields in curved spacetimes \cite{QFT}, by unifying approaches to quantum gravity \cite{strings}, from effective Lagrangians methods \cite{effective}. They are also employed as phenomenological tools to address issues like singularities, early cosmology, astrophysics, and alternatives to the dark matter/energy paradigm [see e.g. \cite{reviews} for some reviews].

The action of Born-Infeld gravity reads
\begin{eqnarray}\label{eq:BI0}
S&=&\frac{1}{\kappa^2\epsilon}\int d^4x \left[\sqrt{-|g_{\mu\nu}+\epsilon R_{\mu\nu}(\Gamma)|}-\lambda \sqrt{-|g_{\mu\nu}|}\right] \nonumber \\
&&+S_m[g_{\mu\nu},\psi] \ ,
\end{eqnarray}
where a vertical bar denotes a determinant, $\epsilon=-2 l_{\epsilon}^2$ is a small parameter with dimensions of length squared and assumed to be negative ($l_{\epsilon}^2>0$), the choice of the constant $\lambda=1$ sets an asymptotically flat spacetime and $S_m$ is the matter action. The required smallness of the length squared scale $\epsilon$ comes from a number of sources, including compatibility with solar systems experiments \cite{Willreviews}, physics of neutron stars \cite{Avelinoneutron,Berti} or binary pulsars \cite{Berti}.  A series expansion of the action (\ref{eq:BI0}) in the parameter $\epsilon$ recovers GR with a cosmological constant term to leading order, and a quadratic Lagrangian $\alpha R^2 + \beta R_{\mu\nu}R^{\mu\nu}$ at next-to-leading order.

In Sec. \ref{sec:II}, we presented the equations that theories like this Born-Infeld model satisfy [see Eq.~(\ref{eq:Rmn})]. For static and spherically symmetric scenarios, the metric and the deformation matrix ${\Omega_\mu}^\alpha$ are diagonal, and exact solutions for electrovacuum configurations can be found analytically (see Refs.~\cite{or,ors} for details). In such configurations, the stress-energy tensor acquires a specific algebraic structure, namely,
\begin{equation}\label{eq:Tmn-generic}
{T^\mu}_\nu=\left(\begin{array}{cc} T_+ \hat I_{2\times2} & \hat O \\ \hat O & T_- \hat I_{2\times2} \end{array}\right) \ .
\end{equation}
Here we will use those solutions to interpret them from the prism of GR, i.e., using them we compute their associated Einstein tensor and interpret their corresponding right-hand side as an effective stress-energy tensor. The properties of the latter will be discussed.

\subsection{Wormhole geometry}

The line element for these electrically charged objects (i.e. described by a standard electrostatic Maxwell field, $\nabla_{\mu}F^{\mu\nu}=0$, with the only non-vanishing component being $F^{tx}$) has been derived in \cite{ors} and takes the form
\begin{equation}\label{eq:ds2}
ds^2=-A(x)dt^2+\frac{1}{A(x)\Omega_{+}^2}dx^2+(r_cz)^2(x)d\Omega^2 \ ,
\end{equation}
with the following definitions: $z\equiv r/r_c$ is a dimensionless radial function with $r_c=\sqrt{r_ql_{\epsilon}}$, and $r_q^2=2G_N q^2$ is a charge scale (where $q$ is the electric charge that emerges from integration of Maxwell equations, i.e., $F^{tx}=q/(r^2 \sqrt{-g_{tt}g_{rr}})$) parametrizing the solutions, the objects $\Omega_{\pm}=1\pm 1/z^4$. The function $A(x)$ is given by
\begin{equation}\label{eq:A}
A(x)= \frac{1}{\Omega_+}\left[1-\frac{(1+\delta_1 G(z))}{\delta_2 z \Omega_{-}^{1/2}}\right] \ ,
\end{equation}
and the two constants parameterizing the metric are
\begin{equation}
\delta_1= \frac{1}{2r_S}\sqrt{\frac{r_q^3}{l_\epsilon}} ,  \qquad
\delta_2= \frac{r_c}{r_S},
\end{equation}
where $r_{S}=2G_N M_0$ is the Schwarzschild radius.

Finally, the function $G(z)$, which satisfies the equation
\begin{equation}
\frac{dG}{dz}=\frac{\Omega_{+}}{z^2 \Omega_{-}^{1/2}} \ ,
\end{equation}
can be analytically integrated and expressed as an infinite power series of the form
\begin{equation}
G(z)=-\frac{1}{\delta_c}+\frac{1}{2}\sqrt{z^4-1}\left[f_{3/4}(z)+f_{7/4}(z)\right] \ ,
\end{equation}
where $f_\lambda(z)={_2}F_1 [\frac{1}{2},\lambda,\frac{3}{2},1-z^4]$ is a hypergeometric function, and $\delta_c\approx 0.572069$ is an integration constant necessary to get the correct asymptotic behavior of GR. Since the radial function $r(x)$ in Eq. (\ref{eq:ds2}) is defined by
\begin{equation} \label{eq:dxdzBI}
x^2= \Omega_{-} r^2  \quad \rightarrow  \quad  \frac{dx}{dz}=\pm\frac{\Omega_{+}}{\Omega_{-}^{1/2}} \ ,
\end{equation}
we will be interested in analyzing the behaviour in two particular limits: far from ($z \gg 1$) and near the region ($z \approx 1$), which respectively means $r \gg r_{c}$ and $r \approx r_{c}$. Moreover, the radial function $r(x)$ defined above can be explicitly written as
\begin{equation} \label{eq:WHc}
r^2(x)= \frac{x^2+\sqrt{x^4+4r_c^4}}{2} \,
\end{equation}
which is a key element in characterizing these solutions.

\begin{figure}[h]
\centering
\includegraphics[width=0.45\textwidth]{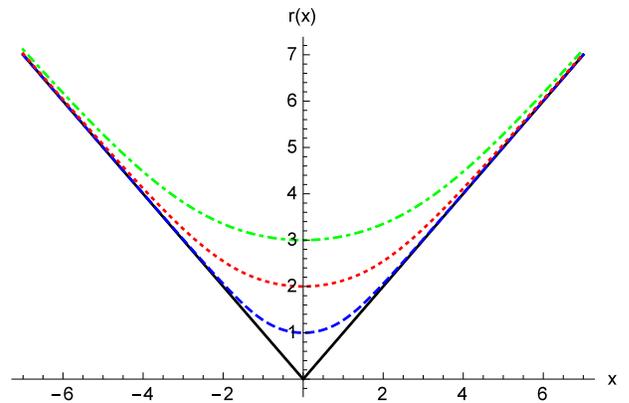}
\caption{Representation of the radial function $r(x)$ (for $r_c=1,2,3$ dashed blue, dotted red and dotted-dashed green, respectively), measured in units of $r_c$. The wormhole throat is located at $x=0$, where the area of the $2$-spheres reaches a minimum and bounces off to another region of spacetime. Note that a few $r_c$ units away from the wormhole throat the radial functions quickly converges to the behaviour of the standard Reissner-Nordstr\"om solution, $r \approx \vert x \vert$ (straight solid black line). \label{fig:1}}
\end{figure}

The geometry encoded in (\ref{eq:ds2}) describes a spacetime which resembles the Reissner-Nordstr\"om (RN) solution of the Einstein-Maxwell field equations of GR as long as $r \gg r_c$. This follows from the expansion of $G(z) \simeq -1/z$ for $z \gg 1$, in such a way that $\Omega_{\pm} \simeq 1$ and
\begin{equation}
A(x) \approx 1-\frac{r_S}{r} + \frac{r_q^2}{2r^2} \ .
\end{equation}
It should be noted that the smallness of the corrections induced by the $\Omega_\pm$ functions implies that the standard GR description is valid all the way down to $r \gtrsim 2r_c$, with noticeable  deviations with respect to the RN solution found only within the region $r \lesssim 2r_c$. The metric function at $r=r_{c}$ behaves as
\begin{eqnarray}\label{eq:A_expansion}
\lim_{r\to r_c} A(x)&\approx& \frac{N_q}{4N_c^\epsilon}\frac{\left(\delta _1-\delta _c\right) }{\delta _1 \delta _c }\sqrt{\frac{r_c}{ r-r_c} }+\frac{1}{2}\left(1-\frac{N_q}{N_c^\epsilon}\right)\nonumber \\&+&O\left(\sqrt{r-r_c}\right) \ .
\end{eqnarray}
We have defined the number of charges as $N_q=q/e$, where $e$ is the proton charge, and introduced the constant $N_c^\epsilon\equiv N_c l_\epsilon/l_P$, where $l_{P}=\sqrt{\hbar G/c^3}$ is the Planck length and $N_c \equiv \sqrt{2/\alpha_{em}} \approx 16.55$ is a constant (with $\alpha_{em}$ the fine structure constant), which will play an important role later. This expansion manifests a strong deviation with respect to the GR behavior. In this region, one finds that the radial function $r(x)$ defined by Eq. (\ref{eq:WHc}) cannot become smaller than $r_c$ (corresponding to $x=0$), where it bounces off to another region of spacetime [see Fig.~\ref{fig:1}]. Thus, the innermost region of these electrically charged objects, which in GR ends up in a point-like singularity at $r=0$, now becomes replaced by a wormhole structure for all values of the charge-to-mass ratio $\delta_1$, with a spherical throat of area $4\pi r_c^2$. When the charge vanishes, $q=0$, the wormhole throat closes and the geometry (\ref{eq:ds2}) describes a standard Schwarzschild black hole, so that Born-Infeld gravity (\ref{eq:BI0}) reduces to GR in vacuum.

Whether the wormhole is covered by a horizon or not depends on the charge-to-mass ratio $\delta_1$, yielding three different structures \cite{ors,or}
\begin{itemize}
  \item $\delta_1>\delta_c$: This is essentially similar to the standard Reissner-Nordstr\"om solution of GR, in that there may be two, one (degenerate) or no horizons on each side of the wormhole.
  \item $\delta_1<\delta_c$: A single (non-degenerate) horizon always exists on each side of the wormhole, resembling the Schwarzschild black hole.
  \item $\delta_1=\delta_c$: This configuration exhibits a single horizon (on each side) which disappears if $N_q<N_c^\epsilon$. For these configurations the metric at $r=r_c$ is finite (i.e. Minkowskian-like).
\end{itemize}
It should be noted that for macroscopic objects ($N_q\gg N_c^\epsilon$) the location of the event horizon is essentially the same as in the Reissner-Nordstr\"om solution of GR, $r_h \approx (r_S+\sqrt{r_S^2-r_q^2})/2$. Only for microscopic solutions ($N_q\sim N_c^\epsilon$)  do we find the qualitative differences of the above classification.

\subsection{Energy conditions}

Let us now assume that the line element (\ref{eq:ds2}) is a solution of Einstein's equations coupled to a certain fluid: $G_{\mu\nu}(g)=\kappa^2\tau_{\mu\nu}$, recalling that $\tau_{\mu\nu}$ is the effective stress-energy tensor. As emphasized above, in such a reinterpretation of the theory, it should not come as a surprise the existence of matter-energy sources in which $T_{\mu\nu}$ satisfies the energy conditions, while $\tau_{\mu\nu}$ does not \cite{Capozziello:2013vna}. Let us apply such a procedure to the spacetime metric (\ref{eq:ds2}).

The components of the effective stress-energy tensor can be written as
\begin{equation}\label{eq:Tmunu}
\tau^{\mu}{}_{\nu}={\rm diag}(-\rho_{\rm eff},-\tau_{\rm eff},p_{\rm eff},p_{\rm eff})\,,
\end{equation}
where $\rho_{\rm eff}$ is the effective energy density, $\tau_{\rm eff}$ is the effective radial tension, and $p_{\rm eff}$ are the effective transverse pressures. Note that this carries the implicit assumption that the effective stress-energy tensor can be cast in a diagonal form, which is reasonable (and can be explicitly verified) given the fact that the extra matter-energy sources appearing in the corresponding field equations are functions of the (diagonal) stress-energy tensor of the electromagnetic field (see Eq. (\ref{eq:Tmn_MOD}) as an example).

In order to follow the standard representation of the wormhole geometry for a static spherically symmetric spacetime \cite{mortho88, visser}, we first use the change of coordinates $dx/dz=\Omega_{+}/\Omega_{-}^{1/2}$ in Eq. (\ref{eq:dxdzBI}) to rewrite the line element (\ref{eq:ds2}) in a more canonical form as
\begin{equation}\label{eq:ds2b}
ds^2=-A(x)dt^2+\frac{1}{A(x)\Omega_{-}}dz^2+z^2(x)d\Omega^2 \ .
\end{equation}
In the standard representation of wormhole geometries the radial coordinate $z^2=z^2(x)$ is assumed to bounce off at $x=0$. In the case considered here, this condition is not imposed but, rather, it is naturally achieved by virtue of the geometry described in the previous section and, in particular, by Eq.~(\ref{eq:WHc}), which arises from the resolution of the corresponding field equations. Note that the use of $z(x)$ as a radial coordinate is subjected to the important restriction that this is only valid in those regions where $z(x)$ is a monotonic function.

Comparing the line element (\ref{eq:ds2b}) with that of the standard wormhole one, Eq. (\ref{eq:whm2}), one simply needs to introduce the identifications
\begin{eqnarray}
\tilde{\Phi}&=&\frac{1}{2} \log (A(x)) \label{eq:PhiA} \ , \\
\tilde{b}(z)&=&z(1-A(x)\Omega_{-}) \label{eq:bA} \ ,
\end{eqnarray}
where we have defined dimensionless variables $\tilde{b}(z)=b(z)/r_c$ and $\tilde{\Phi}(z)=\Phi(z)/r_c$, for consistency in the notation. Now, one just needs to replace the stress-energy tensor (\ref{eq:Tmunu}) into the Einstein field equations to obtain
\begin{eqnarray}
\tilde{\rho}&\equiv&r_c^2 \rho_{\rm eff}=\frac{d\tilde{b}/dz}{z^2} \label{eq:rho} \ , \\
\tilde{\tau}&\equiv&r_c^2 \tau_{\rm eff}= \frac{\tilde{b}(z)}{z^3}-2\left(1-\frac{\tilde{b}(z)}{z} \right) \frac{d\tilde{\Phi}/dz}{z} \label{eq:tau} \ ,\\
\tilde{p}&\equiv&r_c^2p_{\rm eff}=\left(1-\frac{\tilde{b}(z)}{z} \right) \left[\frac{d^2\tilde{\Phi}}{dz^2} + \frac{d\tilde{\Phi}}{dz} \left(\frac{d\tilde{\Phi}}{dz} + \frac{1}{z} \right) \right] \nonumber \\
 &&- \frac{1}{2z^2} \left( \frac{d\tilde{b}}{dz} z - \tilde{b}(z) \right) \left( \frac{d\tilde{\Phi}}{dz} + \frac{1}{z}   \right) \label{eq:p} \ ,
\end{eqnarray}
and the framework is ready to study the matter source threading these wormholes in the context of GR.

Among the spectrum of solutions of our theory, let us just focus on \emph{traversable} wormholes, which correspond to i) solutions with $\delta_1=\delta_c$ and $N_q<N_c^\epsilon$, and ii) solutions with $\delta_1>\delta_c$ and no horizons (naked configurations).

This way, we insert the spacetime metric (\ref{eq:ds2}) into the set of equations (\ref{eq:rho})-(\ref{eq:p}) with the definitions above, and solve them to obtain the components of the effective stress-energy tensor (\ref{eq:Tmunu}) defining the matter source. One then obtains involved expressions for all such components but, nonetheless, we are just interested in exact expressions in the far limit, $r\gg r_c$ (equivalently, $z \gg 1$ or $x \gg 0$), and close to the center, $r=r_c$ (equivalently, $z=1$ or $x=0$). At far distances, one gets
\begin{equation} \label{eq:limit}
\tilde{\rho}(r\gg r_c)=\tilde{\tau}(r\gg r_c)=\tilde{p}(r\gg r_c) \simeq \frac{N_q}{N_c^\epsilon x^4} \ .
\end{equation}
Reverting the notation, this is just the familiar result
\begin{equation}
\tilde{\rho}(r\gg r_c)=\tilde{\tau}(r\gg r_c)=\tilde{p}(r\gg r_c) \simeq \frac{q^2}{r^4} \ ,
\end{equation}
which are the energy, tension and pressure of the Maxwell field supporting the standard Reissner-Nordstr\"om solution, in agreement with the recovery of this solution in the far limit. This implies that for these solutions no violation of the energy conditions occur for far distances.

\begin{figure}[h]
\centering
\includegraphics[width=0.45\textwidth]{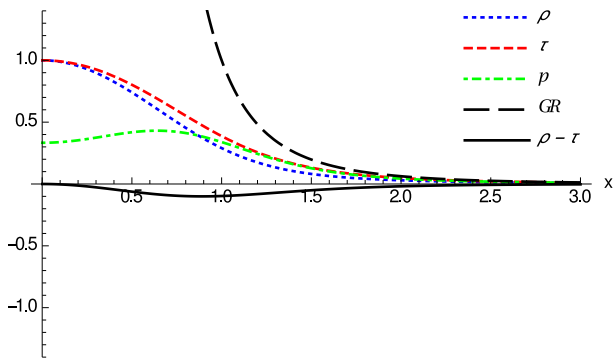}
\caption{Graphical representation of the components $\tilde{\rho}$ (blue dotted), $\tilde{\tau}$ (red thin dashed) and $\tilde{p}$ (green dashed-dotted) of the stress-energy tensor (\ref{eq:Tmunu}) as a function of $x$ for regular horizonless solutions ($\delta_1=\delta_c$) with $N_q=N_c^\epsilon$, which represents the transition case where the horizon disappears. At large distances all these components reduce to those of the Reissner-Nordstr\"om case (black thick dashed, upper curve in this figure), $N_q/(N_c^\epsilon x^4)$ [see Eq.~(\ref{eq:limit})]. The solid black line (lower curve in this figure) represents $ \tilde{\rho}-\tilde{\tau}<0$, which means that the NEC is violated near the wormhole throat.  \label{fig:2}}
\end{figure}

\begin{figure}[h]
\centering
\includegraphics[width=0.45\textwidth]{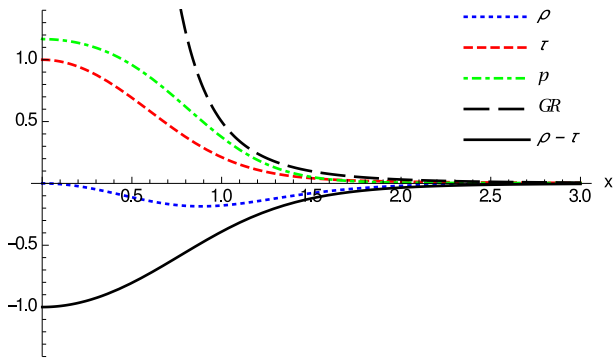}
\caption{Same notation and curves as in Fig.~\ref{fig:2}, for $\delta_1=\delta_c$ now with $N_q=N_c^\epsilon/2$ (for $l_\epsilon=l_P$). Here the NEC is violated because $ \tilde{\rho}-\tilde{\tau} <0$ (solid black). Note also that though $\tilde{\rho}<0$ near the throat, far away it becomes positive ($\tilde{\rho}\approx q^2/r^4$). \label{fig:3}}
\end{figure}

Close to the spherical surface $r=r_c$ (or $z=1$), where $x=0$ (see Eq.~(\ref{eq:WHc})), the expansion of the components of the stress-energy tensor yields the result
\begin{eqnarray}
\lim_{x\to 0}\tilde{\rho} &\approx& -\frac{N_q(\delta_1-\delta_c)}{N_c^\epsilon \delta_c \delta_1 |x|}  + \left( \frac{2N_q}{N_c^\epsilon}-1 \right) + \mathcal{O}(x^2) \label{eq:rhoc} \ , \\
\lim_{x\to 0}\tilde{\tau} &\approx& \frac{N_q(\delta_1-\delta_c)}{N_c^\epsilon \delta_c \delta_1 |x|} + 1 -\frac{2N_q(\delta_1-\delta_c)}{N_c^\epsilon \delta_c \delta_1} |x| +\mathcal{O}(x^2) \ , \\
\lim_{x\to 0}\tilde{p}    &\approx& \frac{2N_q(\delta_1-\delta_c)}{N_c^\epsilon \delta_c \delta_1 |x|^3} - \frac{N_q (\delta_1 - \delta_c)}{N_c^\epsilon \delta_c \delta_1 |x|}  \\
&&+ \left(2- \frac{5N_q}{3N_c^\epsilon} \right)  + \mathcal{O}(x^2) \nonumber \ .
\end{eqnarray}
To further interpret these results let us bring to our previous discussion the standard classification of the energy conditions. From among the four (null, weak, strong and dominant energy conditions) we are mostly interested in the NEC, because its violation implies the violation of all the others. Since the matter stress-energy tensor has the form given by Eq.~(\ref{SETdef}), the NEC implies that
\begin{equation}
\tilde{\rho} + \tilde{p}_j \geq 0, \quad \forall j=1,2,3 \,,
\end{equation}
where $\tilde{p}_1=-\tilde{\tau}$ and $\tilde{p}_2=\tilde{p}_3=\tilde{p}$. To study this condition, we observe that around the wormhole throat
\begin{equation} \label{eq:trho}
\lim_{x\to 0} (\tilde{\rho}-\tilde{\tau}) \simeq -\frac{2N_q (\delta_1-\delta_c)}{N_c^\epsilon\delta_c\delta_1 |x|} + 2\left(\frac{N_q}{N_c^\epsilon}-1  \right) + \mathcal{O}(x) \ ,
\end{equation}
and thus it is clear that for horizonless remnants, $\delta_1=\delta_c$ and $N_q < N_c^\epsilon$, the first term is zero and the leading order term becomes negative at the wormhole throat, $\lim_{x\to 0} (\tilde{\rho}-\tilde{\tau})<0$. As depicted in Fig.~\ref{fig:2}, this means that the NEC is violated in a small region around $x=0$. The fact that only exotic matter can support these geometries is in complete agreement with the standard wisdom of wormholes within GR. Note that whether the energy density becomes negative, $\tilde{\rho}<0$, or not, depends on additional constrains upon the number of charges $N_q$, as can be seen from the comparison between Figs. \ref{fig:2} and \ref{fig:3}.
\begin{figure}[h]
\centering
\includegraphics[width=0.45\textwidth]{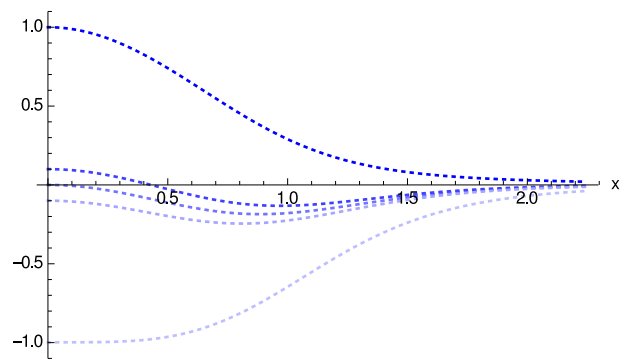}
\caption{Representation of $\tilde{\rho}$ for different values of $N_q$. The upper curve corresponds to $N_q=N^\epsilon_c$ and the lower to $N_q\to 0$. The middle curves represent, respectively, $N_q>N^\epsilon_c/2$, $N_q=N^\epsilon_c/2$ and $N_q<N^\epsilon_c/2$ and illustrate how the energy density at the throat changes its sign as the number of charges decreases (in a continuum way). \label{fig:7}}
\end{figure}

In Fig.~\ref{fig:7} we illustrate how the energy density near the wormhole throat changes sign as the number of charges decreases from its critical value $N_c^\epsilon$. It should be noted that this figure represents the (dimensionless) energy density $\tilde{\rho}$, which hides a certain dependence on the number of charges $N_q$. The effect of this dependence is relevant, as $\rho_{\rm eff}\sim \tilde{\rho}/N_q$, thus implying that the energy density becomes very negative as $N_q\to 0$. Note also that the $x$-axis should also be rescaled to account for the dependence on $N_q$, which leads to $x\to x N_q^{1/2}$. Thus, the region where the energy density becomes very negative is also restricted to a tiny region near the throat. In the limit $N_q\to 0$ then one finds that this region is of zero size, which is consistent with the fact that without electric charge there is no wormhole and the Schwarzschild solution is recovered.

\begin{figure}[h]
\centering
\includegraphics[width=0.45\textwidth]{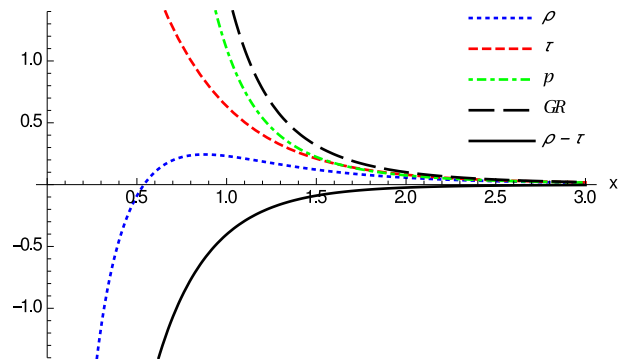}
\caption{Same notation and curves as in Fig.~\ref{fig:2}, now with $\delta_1=\delta_c +0.3$ and $N_q=N_c^\epsilon +10$, corresponding to Reissner-Nordstr\"om-like configurations without horizons (naked). Note that the NEC is violated because $ \tilde{\rho}-\tilde{\tau}<0$ (solid black line) around the wormhole throat $x=0$. Here $\tilde{\rho}$ and $\tilde{\tau}$ diverge as $1/|x|$ when $|x|\to 0$. \label{fig:4}}
\end{figure}

For configurations with $\delta_1>\delta_c$ the existence or absence of horizons depends on the charge-to-mass ratio, which should be large (more charge than mass in certain units). Nonetheless, it is clear from Eqs.~(\ref{eq:rhoc}) and (\ref{eq:trho}) that the NEC will be violated because $\lim_{x\to 0}(\tilde{\rho}-\tilde{\tau}) <0$ (and one always has $\lim_{x\to 0}\tilde{\rho}<0$). When the GR solutions are a good approximation, the absence of horizons translates into the condition $2r_q^2/r_S^2>1$ which, knowing that $\delta_1=(N_q/N_c^\epsilon) \delta_2$, can be written as $(\delta_1/\delta_c)\gtrsim \sqrt{N_q/N_c^\epsilon}$. In Fig. \ref{fig:4}, we plot a particular case with $\delta_1=\delta_c+0.3$ and $N_q=N_c^\epsilon +10$, which satisfies the above no-horizons constraint. There it is seen that the NEC is violated around the throat by an infinite amount, which is in contrast with the finite violation achieved in the $\delta_1=\delta_c$ cases. By completeness, we also point out that the flaring-out condition can be numerically checked to be satisfied both in the $\delta_1>\delta_c$  and $\delta_1=\delta_c$ cases above.

From our analysis it is clear that the emergence of the wormhole structure from the prism of GR is related to the deviations that the stress-energy tensor suffers with respect to that of a Maxwell field as the innermost region $r=r_c$ is approached. This is so because the symmetry satisfied  by the stress-energy tensor characterizing electromagnetic fields, namely, ${T_t}^t={T_r}^r$ and ${T_\theta}^{\theta}={T_\varphi}^{\varphi}$ cannot support wormhole solutions in the context of GR, even for non-linear electromagnetism \cite{Arellano}. Thus, one is forced to consider exotic sources of matter to generate the wormhole. Alternatively, one can reconsider this scenario and interpret these wormholes as emerging out of modified gravity effects, in which case electromagnetic fields (either Maxwell or non-linear \cite{or14}) with ${T_t}^t={T_r}^r$ and ${T_\theta}^{\theta}={T_\varphi}^{\varphi}$ but ${\tau_t}^t\neq{\tau_r}^r$ and ${\tau_\theta}^{\theta}\neq{\tau_\varphi}^{\varphi}$  can naturally sustain them without violation of the energy conditions.

Note that in all cases the typical size of the region where these violations of the energy conditions occur is determined by the scale $r_c=\sqrt{r_q l_{\epsilon}}=l_\epsilon \sqrt{2N_q/N_c^\epsilon}$ (observe that the $x$-scale and the vertical scale in Figs. \ref{fig:2}--\ref{fig:3} are both measured in units of $r_c$). This implies that violations of the energy conditions are restricted to a region of order $\sqrt{l_Pl_\epsilon}$ which may grow with the charge as $N_q^{1/2}$ (recall that $N_c^\epsilon\approx 16.55 l_\epsilon/l_P$). Thus, unless $l_\epsilon$ be much greater than $l_P$ and/or one considers huge amounts of charge, the size of this region will typically be  very small.

Let us stress that though curvature divergences may arise at the wormhole throat $r_c$, these wormholes are geodesically complete for all spectra of mass and charge and physical observers can safely traverse them \cite{geo1}. This should not come as a surprise since the same feature is quite common in the thin-shell approach discussed above.

\section{$f(R)$ wormholes} \label{sec:V}

For the discussion of this type of wormholes, we restrict ourselves to quadratic $f(R)$ models of the form \footnote{This model was originally introduced by Starobinsky \cite{Starobinsky}. In this section we focus on the choice $\lambda >0$ for which wormhole solutions have been obtained in the literature, though wormholes for $\lambda<0$ are also possible \cite{BOR2017}.}
\be \label{eq:action}
S=\frac{1}{2\kappa^2} \int d^4x \sqrt{-g} (R-\lambda R^2) + S_m(g_{\mu\nu},\psi_m) \ ,
\en
where $\lambda$ is a constant with dimensions of length squared. The matter sector is represented by an anisotropic fluid whose stress-energy tensor reads
\begin{equation} \label{eq:Tmunua}
T^{\mu}{}_{\nu}= {\rm diag}(-\rho,-\rho,\alpha \rho, \alpha \rho) \ ,
\end{equation}
where $0 \leq \alpha \leq 1$ is necessary to satisfy the energy conditions. This type of fluids has been recently considered in order to investigate the existence of wormhole solutions in several extensions of GR \cite{bi}. Note that the case $\alpha=1$ corresponds to a standard Maxwell field, which has a vanishing trace for ${T_\mu}^{\nu}$. Due to the fact that only matter sources with a nonzero trace can excite the dynamics of Palatini $f(R)$ gravity, we shall be concerned here with the cases $\alpha \neq 1$. Let us also point out that some models of non-linear electrodynamics have the same structure (\ref{eq:Tmunua}) in their stress-energy tensor.

\subsection{Wormhole geometry}

The solutions of this gravity-matter system are characterized by the following line element, suitably adapted to our problem \cite{or15}:
\begin{equation} \label{eq:geomf}
ds^2=-A(z)dt^2 + \frac{1}{A(z)f_R^3} \left(1+\frac{\alpha}{z^{2+2\alpha}}\right)^2 dz^2 + z^2(x)d\Omega^2 \ ,
\end{equation}
where $z=r/r_c$ is now defined in terms of $r_c^{2+2\alpha}=1-(4\lambda) \kappa^2 (1-\alpha)C$  ($C$ is a dimensional integration constant such that the energy density reads $\rho(x)=C/r(x)^{2+2\alpha}$) and the metric function can be obtained in closed analytical form as
\begin{equation} \label{eq:A-fdef}
A(z)=\left(1-\frac{1+\delta_1 G_\alpha(z)}{\delta_2 z f_R^{1/2}}\right)f^{-1}_R \ ,
\end{equation}
with $f_R=1-z^{-(2+2\alpha)}$, where we have defined the following constants: $\delta_1 = r_c^3/(8\lambda M_0)$ and $\delta_2=r_c/(2M_0)$. The function $G_\alpha(z) $ is given by
\begin{eqnarray}
\label{eq:GfR}
&&G_\alpha(z)  = \frac{z^{-4 \alpha -1}}{2 (\alpha -1) (2 \alpha -1)} \times
	\nonumber  \\
&& \times \left[
\frac{z^{2 \alpha +2} \sqrt{1-z^{-2 (\alpha +1)}} \left(2 \alpha ^2+\alpha +2 z^{2 \alpha +2}-3\right)}{\left(z^{2 \alpha +2}-1\right) 2 (\alpha -1) (2 \alpha -1)} \right.
	\nonumber  \\
&& - \left. \frac{8 \alpha  \left(\alpha ^2-1\right) \, _2F_1\left(\frac{1}{2},\frac{4 \alpha +1}{2 \alpha +2};\frac{6 \alpha +3}{2 \alpha +2};z^{-2 (\alpha +1)}\right)}{\left(4 \alpha +1\right)} \right]
\ ,
\end{eqnarray}
where $_2F_1\left(a, b, c; y\right)$ is a hypergeometric function and $M_0$ is the Schwarzschild mass. Note that the expression for $G_\alpha(z)$ is valid as long as $\alpha \neq 1$ and $\alpha \neq 1/2$, in which cases different expressions are found [see \cite{or15} for details].

\begin{figure}[h]
\centering
\includegraphics[width=0.45\textwidth]{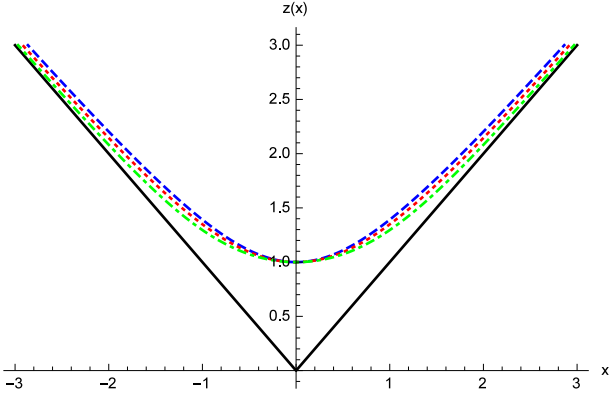}
\caption{Behaviour of the radial function $z(x)$ in the $f(R)$ gravity case, obtained from Eq.~(\ref{eq:WHth}), for the cases $\alpha=1/10,1/3,3/4$ (dashed blue, dotted red and dotted-dashed green, respectively)). At $x \gg 1$ the function $z^2(x) \simeq x^2$ and the GR behaviour is restored (straight solid black line). Observe the bouncing behaviour of all these curves at $x=0$.\label{fig:5}}
\end{figure}

Like in the previous case, the wormhole structure is manifest from the behaviour of the radial function $r(x)$, which satisfies the equation
\begin{equation} \label{eq:WHth}
\frac{dz}{dx}= \frac{f_R^{1/2}}{\left(1+\frac{\alpha}{z^{2+2\alpha}} \right)}  \quad  \Rightarrow  \quad  x=z\sqrt{1-\frac{1}{z^{2+2\alpha}}}\,,
\end{equation}
which could be used to rewrite the geometry (\ref{eq:geomf}) with $x$ as the radial coordinate, similarly as in Eq. (\ref{eq:ds2}) of the Born-Infeld gravity wormholes. Far from the center, $z \gg 1$, one recovers $r \simeq x$, in agreement with the fact that the corrections to GR vanish there. Close to the center $z \approx 1$ (where $r=r_c$ or $x=0$), one finds that the radial function reaches a minimum and bounces off, similarly to what was found in the previous section, though in this case a closed expression for $z(x)$ is not easy to obtain for generic $\alpha$. Nonetheless, in Fig.~\ref{fig:5} we have numerically integrated the relation (\ref{eq:WHth}), where we see the expected bouncing behavior.

Following the definitions introduced above, the shape function $\tilde{b}(z)$ characterizing the wormhole is now written as
\begin{equation}
\tilde{b}(z)=z \left(1-\frac{A(z)f_R^3}{\left(1+\frac{\alpha}{z^{2+2\alpha}} \right)^2} \right) \ ,
\end{equation}
while the redshift function $\tilde{\Phi}(z)$ is still formally given by Eq.~(\ref{eq:PhiA}), now with the set of definitions (\ref{eq:A-fdef})--(\ref{eq:GfR}).

For arbitrary $\alpha$, a series expansions both at $z \rightarrow \infty$ and as $z \simeq 1$ yield long expressions, which are of little use for our purposes. Therefore we shall consider a particular case, say $\alpha=3/4$, to illustrate the analysis. First, we note that the asymptotic behavior of the metric becomes
\begin{equation}
A(z) \approx 1-\frac{\delta_2}{z} + \frac{8\delta_1}{\delta_2z^{3/2}} + \mathcal{O}\left(\frac{1}{z^{7/2}}\right) \label{eq:Aasym} \ ,
\end{equation}
whose fall off is governed by the standard Schwarzschild mass term $\delta_2/z$, and is asymptotically flat. Note that the charge term contribution has been softened, from the $\sim 1/z^2$ contribution of the Reissner-Nordstr\"om spacetime, to the $\sim 1/z^{3/2}$ of the present case. Indeed, this is a general property of all these models with $1/2<\alpha<1$, whereas those with $0<\alpha<1/2$ are still asymptotically flat, but governed by the ``charge" term instead of the Schwarzschild mass term. This is in agreement with the fact that the stress-energy tensor of the fluid given by Eq.~(\ref{eq:Tmunua}) can also be reproduced by non-linear theories of electrodynamics with $L_m=X^{\alpha}$, where $X=-\frac{1}{2} F_{\mu\nu}F^{\mu\nu}$, and those models with $0<\alpha<1/2$ have been found to be asymptotically flat but not Schwarzschild-like (\emph{anomalous} models \cite{dr2}). Note that, as opposed to the Maxwell case, $L_M=X$, the trace of such models is non-vanishing, which allows to excite the non-linear dynamics driven by the $f(R)$ gravity [since in these models $R=-\kappa^2 T$].

\subsection{Energy conditions}

Using Eq.~(\ref{eq:Aasym}) above, one can get the asymptotic behavior of the energy density and the pressures, namely
\begin{eqnarray}
\tilde{\rho}(z \gg 1)&\approx& 4\frac{\delta_1}{\delta_2}\frac{1}{z^{7/2}} - \frac{35}{4}\frac{1}{z^{11/2}} + \mathcal{O}\left(\frac{1}{z^{13/2}} \right) \ , \\
\tilde{\tau}(z \gg 1)&\approx& 4\frac{\delta_1}{\delta_2}\frac{1}{z^{7/2}} +  \,\,  7\frac{1}{z^{11/2}} + \mathcal{O}\left(\frac{1}{z^{13/2}} \right) \ , \\
\tilde{p}(z \gg 1) &\approx&  3\frac{\delta_1}{\delta_2}\frac{1}{z^{7/2}} + \frac{49}{4}\frac{1}{z^{11/2}} + \mathcal{O}\left(\frac{1}{z^{13/2}} \right) \ .
\end{eqnarray}
To first order we recover the GR expressions, which satisfy the energy conditions. This contrasts with the higher-order corrections generated by the $f(R)$ dynamics, which violate the NEC as $\tilde{\rho}(z \gg 1)-\tilde{\tau}(z \gg 1)<0$. From our definitions, one verifies that such corrections are highly suppressed by a factor $\sim (\lambda \kappa^2 C)^{11/7}/r^{11/2}$. Thus, for sufficiently small values of the length squared scale $\lambda$ such asymptotic violations of the energy conditions would lie far beyond experimental reach.

\begin{figure}[h]
\centering
\includegraphics[width=0.45\textwidth]{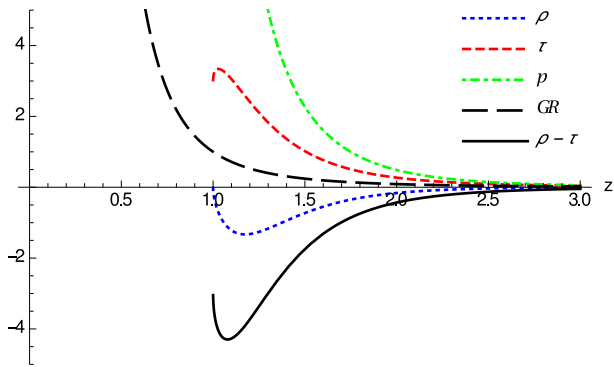}
\caption{Same notation and curves as in Fig.~\ref{fig:2}, now for the $f(R)$ gravity case with $\alpha=3/4$ and parameters $\delta_2=4\delta_1$ and $\delta_1=10$, for which no horizons are present. Note that the NEC is violated as $ \lim_{x\to 0}(\tilde{\rho}-\tilde{\tau})<0$ (solid black) around the wormhole throat $z=1$.  Note that though $\tilde{\rho}<0$ near the throat, far away it becomes positive recovering the GR limit. \label{fig:6}}
\end{figure}

In the opposite limit, near the wormhole (now we expand around the wormhole throat located at $z=1$ instead of $x$ since, as mentioned above, it is not possible to obtain a closed expression $z=z(x)$), we find
\begin{eqnarray}
\lim_{z \to 1} \tilde{\rho} &\approx& 1-4\frac{\delta_1}{\delta_2} + \frac{3\left(8-\frac{21\sqrt{\pi} \Gamma[\frac{15}{7}]\delta_1}{\Gamma[\frac{23}{14}]} \right)}{2\sqrt{14}\delta_2} (z-1)^{1/2} \nonumber \\
&&+ \mathcal{O}(z-1) \,, \label{eq:fr1} \\
\lim_{z \to 1} \tilde{\tau} &\approx& 1+8\frac{\delta_1}{\delta_2} - \frac{3\left(8-\frac{21\sqrt{\pi} \Gamma[\frac{15}{7}]\delta_1}{\Gamma[\frac{23}{14}]} \right)}{2\sqrt{14}\delta_2} (z-1)^{1/2} \nonumber \\
&&+ \mathcal{O}(z-1) \,,  \label{eq:fr2}  \\
\lim_{z \to 1} \tilde{p} &\approx& 6\frac{\delta_1}{\delta_2}\frac{1}{(z-1)} - \frac{3\left(8-\frac{21\sqrt{\pi} \Gamma[\frac{15}{7}]\delta_1}{\Gamma[\frac{23}{14}]} \right)}{4\sqrt{14}\delta_2} \frac{1}{(z-1)^{1/2}}  \nonumber \\
&&+ \mathcal{O}(z-1)  \,, \label{eq:fr3}
\end{eqnarray}
where $\Gamma[a]$ is Euler's gamma function. From these expressions it is immediately seen that $\lim_{z\to 1}(\tilde{\rho}-\tilde{\tau})<0$ and thus the NEC is violated regardless of the values of the constants $\delta_1$ and $\delta_2$ characterizing the solutions. In addition, when $\delta_1/\delta_2>1/4$ one has that $\lim_{z\to 1}\tilde{\rho}<0$. In Fig.~\ref{fig:6}, the different components of the stress-energy tensor are plotted for the case $\delta_2=4\delta_1$ and $\delta_1=10$, in which one can numerically check the absence of horizons and thus we are dealing with traversable wormholes. In this plot we observe a bounded violation of the NEC near the throat, similarly to what was found in the $\delta_1=\delta_c$ configurations of the Born-Infeld gravity theory studied in the previous section. In addition, one can numerically check that the flaring-out condition is satisfied for these wormhole geometries.

Let us stress that, like in the Born-Infeld gravity case, for the existence of these wormhole solutions it is not enough to have a non-linear electrodynamics source. This follows from the fact that in GR no wormholes supported by this kind of matter source are allowed \cite{Arellano}. Indeed, as follows from Eqs.~(\ref{eq:fr1})--(\ref{eq:fr3}) the \emph{effective} stress-energy tensor that sources our wormholes does not have the symmetries of nonlinear theories of electrodynamics, namely, $\tau_{t}^t \neq \tau_x^x$ and $\tau_{\theta}^{\theta} \neq \tau_{\varphi}^{\varphi}$, due to the presence of $\lambda$-corrections. When $\lambda$ vanishes, the wormhole throat closes and one recovers the GR results. Note that the preceding discussion could be further extended to include other theories of (non-linear) electrodynamics, where wormholes have been recently found \cite{bcor15}.

\section{Conclusion and Discussion} \label{sec:VI}

In this work, we have presented two classes of traversable wormhole  spacetimes which are supported by a single matter source  given in the action from which the geometry is derived and turns out to be well-defined everywhere. These spacetimes are exact solutions of certain extensions of GR including higher-order powers and contractions of the Ricci tensor, and formulated in the Palatini approach. Here we have shown that they can also be interpreted as exact solutions of GR with a modified or effective stress-energy tensor. We stress that these solutions are not constructed as a result of the standard reverse-philosophy in general relativistic wormhole physics, but instead directly arise from well-defined actions. Indeed, these features have been obtained not as a consequence of any designer or engineering process but as the result of a direct derivation from well motivated gravitational actions. The keystone seems to be the Palatini framework where the original theories were formulated.

From the GR viewpoint, we have found that the generalized energy conditions for wormholes coming from Born-Infeld and $f(R)$ gravity are violated at/near the wormhole throat, which is in perfect agreement with the current knowledge of wormholes physics. Although the kind of matter source, needed to sustain a wormhole from this point of view, is quite peculiar, it is confined in a restricted region around the wormhole throat and more important it does not seem to be in contradiction with any fundamental physics principle. However, from the modified gravity side there is a straightforward way to reformulate the standard view by describing a wormhole spacetime actually threaded by ordinary matter but supported by the novel gravitational effects attached to the particular extended gravity and whose size is determined by a fundamental length scale, which in turn defines the size of the violations of the energy conditions from the GR viewpoint. Certainly, the properties or the behavior of matter could be dependent on the theoretical framework adopted. Therefore, it is an overriding issue to keep exploring alternative theories of gravity. The Palatini approach is indeed an encouraging new perspective by making possible to deal with the problem of singularities by replacing them with microscopic wormholes sustained by ordinary matter.

In summary, the wormhole spacetimes presented in this work arise in  a more natural (less artificial) way in Palatini formalism where metric and connection are assumed as independent dynamical variables. This novel approach may bring up new avenues to enlarge our knowledge of wormhole physics, offer new insights for constructing wormhole solutions in the context of GR, and find reasonable scenarios where they might take place.

\section*{Acknowledgments}

C. B. is funded by the National Scientific and Technical Research Council (CONICET). F. S. N. L. acknowledges financial  support of the Funda\c{c}\~{a}o para a Ci\^{e}ncia e Tecnologia (FCT) through an Investigador Research contract, with reference IF/00859/2012, funded by FCT/MCTES (Portugal). G. J. O. is supported by a Ramon y Cajal contract and the Spanish grant FIS2014-57387-C3-1-P from MINECO. This work has also been supported by the i-COOPB20105 grant of the Spanish Research Council (CSIC),  the Consolider Program CPANPHY-1205388, the Severo Ochoa grant SEV-2014-0398, and the CNPq (Brazilian) project No.301137/2014-5. D. R.-G. is funded by the FCT postdoctoral fellowship No.~SFRH/BPD/102958/2014 and the FCT research grant UID/FIS/04434/2013. C. B. and D. R.-G. thank the Department of Physics of the University of Valencia for their hospitality during the initial stage of this work. This article is based upon work from COST Action CA15117, supported by COST (European Cooperation in Science and Technology).



\end{document}